\documentclass{aa}  
\usepackage{graphicx}
\usepackage{txfonts}
\usepackage[colorlinks,citecolor=blue,urlcolor=blue,filecolor=blue,linkcolor=blue]{hyperref}
\usepackage{amsmath}
\usepackage{amssymb}
\usepackage{upgreek}
\usepackage{natbib}
\usepackage{xcolor}

\newcommand{\msun}{\, {\rm M}_\odot}

\begin{document} 
    \title{%Discovery of a diffuse radio ring around the Calvera pulsar}
    Possible discovery of Calvera's supernova remnant}
   
    \author{M.\ Arias\inst{1}
        \and
        A.\ Botteon\inst{1}
        \and
        C.\ G.\ Bassa\inst{2}
        \and
        S.\ van der Jagt\inst{1}
        \and
        R.\ J.\ van Weeren\inst{1}
        \and
        S.~P.\ O'Sullivan\inst{3}
        \and
        Q.\ Bosschaart\inst{1}
        \and
        R.\ S.\ Dullaart\inst{1}
        \and
        M.\ J.\ Hardcastle \inst{4}
        \and
        J.\ W.\ T. Hessels\inst{2,5}
        \and
        T.\ Shimwell\inst{2}
        \and
        M.\ M.\ Slob\inst{1}
        \and
        J.\ A.\ Sturm
        \and
        C.\ Tasse\inst{5,6}
        \and
        N.\ C.\ M.\ A.\ Theijssen\inst{1}
        \and
        J. Vink\inst{7}
    }

    \institute{
        Leiden Observatory, Leiden University, P.O.\ Box 9513, 2300\,RA Leiden, The Netherlands\\
        \email{arias@strw.leidenuniv.nl}  
        \and
        ASTRON Netherlands Institute for Radio Astronomy, Oude Hoogeveensedijk 4, 7991\,PD Dwingeloo, The Netherlands
        \and
        School of Physical Sciences and Centre for Astrophysics \& Relativity, Dublin City University, Glasnevin, D09 W6Y4, Ireland
        \and
        Centre for Astrophysics Research, Department of Physics, Astronomy and Mathematics, University of Hertfordshire, College Lane, Hatfield AL10 9AB, UK
        \and
        GEPI \& USN, Observatoire de Paris, Universit\'e PSL, CNRS, 5 Place Jules Janssen, 92190 Meudon, France \and
        Department of Physics \& Electronics, Rhodes University, PO Box 94, Grahamstown, 6140, South Africa
        \and
        Anton Pannekoek Institute for Astronomy, University of Amsterdam, Science Park 904, 1098\,XH Amsterdam, The Netherlands \\
    }

    \date{Received \today; Accepted }

    \abstract{We report the discovery of a ring of low surface brightness radio emission around the Calvera pulsar, a high Galactic latitude, isolated neutron star, in the LOFAR Two-metre Sky Survey (LoTSS). It is centered at $\alpha=14\mathrm{h}11\mathrm{m}12\fs6$, $\delta=+79\degr23\arcmin15\arcsec$, has inner and outer radii of $14\farcm2$ and $28\farcm4$, and an integrated flux density at 144 MHz of $1.08\pm0.15$\,Jy. The ring center is offset by $4\farcm9$ from the location of the Calvera pulsar. H$\alpha$ observations with the Isaac Newton Telescope show no coincident optical emission, but do show a small ($\sim20$\arcsec) optical structure internal to the ring.  We consider three possible interpretations for the ring: that it is an H~II region, a supernova remnant (SNR), or an Odd Radio Circle (ORC). The positional coincidence of the ring, the pulsar, and an X-ray-emitting non-equilibrium ionisation plasma previously detected, lead us to prefer the SNR interpretation. If the source is indeed a SNR and its association with the Calvera pulsar is confirmed, then Calvera's SNR, or G118.4+37.0, will be one of few SNRs in the Galactic halo.}


    \maketitle
    
\section{Introduction}
\iffalse
Radio surveys have been crucial tools in advancing our understanding of the Universe since the advent of modern astronomy. 
From classical discoveries such as the Galactic background radiation \citep{jansky33}, pulsars \citep{hewish68}, active galactic nuclei \citep{bolton49},
or quasars \citep{hazard63}, to the cosmological implications of the source-count slope of radio luminosities \citep{ryle55}, radio
surveys have touched most fields in the discipline. 
The current generation of deep radio surveys are poised to continue this tradition of profound impact. 

The LOFAR Two-metre Sky Survey \cite[LoTSS,][]{shimwell17} is an ongoing survey of the Northern sky at $120-168$~MHz, with $6\arcsec$ resolution
and $\lesssim100$\,$\upmu$Jy\,beam$^{-1}$ sensitivity\footnote{The LoTSS data are taken with the full configuration of the LOFAR array, including international stations, but only the Dutch configuration of the array is processed. The data are available for reprocessing to $0.3\arcsec$ resolution.}. There have been two LoTSS data releases \citep{shimwell19,shimwell22}
that have enabled a variety of scientific results in fields such as planetary nebulae \citep{hajduk21}, star-planet interaction \citep{callingham21}, radio pulsars \citep{tan18}, Galactic polarisation \citep{erceg22}, supernova remnants \citep{arias19}, galaxy clusters \citep{botteon22}, giant radio galaxies \citep{oei22}, the extragalactic radio background \citep{hardcastle21}, and cosmic magnetic fields \citep{osullivan20,stuardi20,carretti22}. 
\fi

The LOFAR Two-metre Sky Survey \cite[LoTSS,][]{shimwell17} is an ongoing survey of the Northern sky at $120-168$~MHz, with $6\arcsec$ resolution
and $\lesssim100$\,$\upmu$Jy\,beam$^{-1}$ sensitivity\footnote{The LoTSS data are taken with the full configuration of the LOFAR array, including international stations, but only the Dutch configuration of the array is processed. The data are available for reprocessing to $0.3\arcsec$ resolution.}. There have been two LoTSS data releases \citep{shimwell19,shimwell22}
that have enabled a wide variety of scientific results.

While processing proprietary data from LoTSS, we have discovered a ring of diffuse radio emission encompassing the celestial position of the isolated neutron star Calvera (1RXS\,J141256.0+792204). 
It is a degree-scale circular structure at high Galactic latitude ($l = 37\degr$), that may or may not be associated with the Calvera neutron star. 

1RXS\,J141256.0+792204 is an isolated neutron that so far defies classification into the established types of neutron
stars. Initially identified as a candidate thermally emitting isolated neutron star by \citet{rutledge08}, selected on its soft X-ray spectrum and high X-ray to optical flux ratio from the \textit{\textit{ROSAT}} All-Sky Survey \citep{voges99}, it was nicknamed ``Calvera'' due to its similarities with the 7 nearby X-ray dim isolated neutron stars (XDINS), colloquially named the ``Magnificent Seven'' \citep{haberl07}. However, its high high black body temperature set it apart from the XDINS, a distinction further confirmed by the discovery of X-ray pulsations at a 59 ms spin period \citep{zane11}, much shorter than the several second spin periods of the ``Magnificent Seven'', and by a distance estimate of 3.3~kpc \citep{mereghetti21}, much larger than the few hundred parsec distances of the other XDINS. X-ray timing yields dipolar spindown properties ($\dot{E}=6.1\times10^{35}$\,erg\,s$^{-1}$, $\tau_\mathrm{c}=2.9\times10^{5}$\,yr, $B=4.4\times10^{11}$\,G) that are more in line with those of young radio pulsars \citep{halpern13,bogdanov19,mereghetti21}, disfavouring
the classification as a nearby millisecond pulsar or a central compact object (CCO). On the other hand, the absence of radio pulsations \citep{hessels07,zane11} or gamma-ray pulsations \citep{halpern11,halpern13,mereghetti21} are somewhat inconsistent with those of young radio pulsars.

Here, we report the detection of a diffuse radio ring around the Calvera pulsar, and discuss three possible scenarios: that it is an H~II region, that it is a supernova remnant (SNR), possibly associated with Calvera, or that it is an Odd Radio Circle (ORC). 

\begin{figure*}
    \centering
    \includegraphics[width=\textwidth]{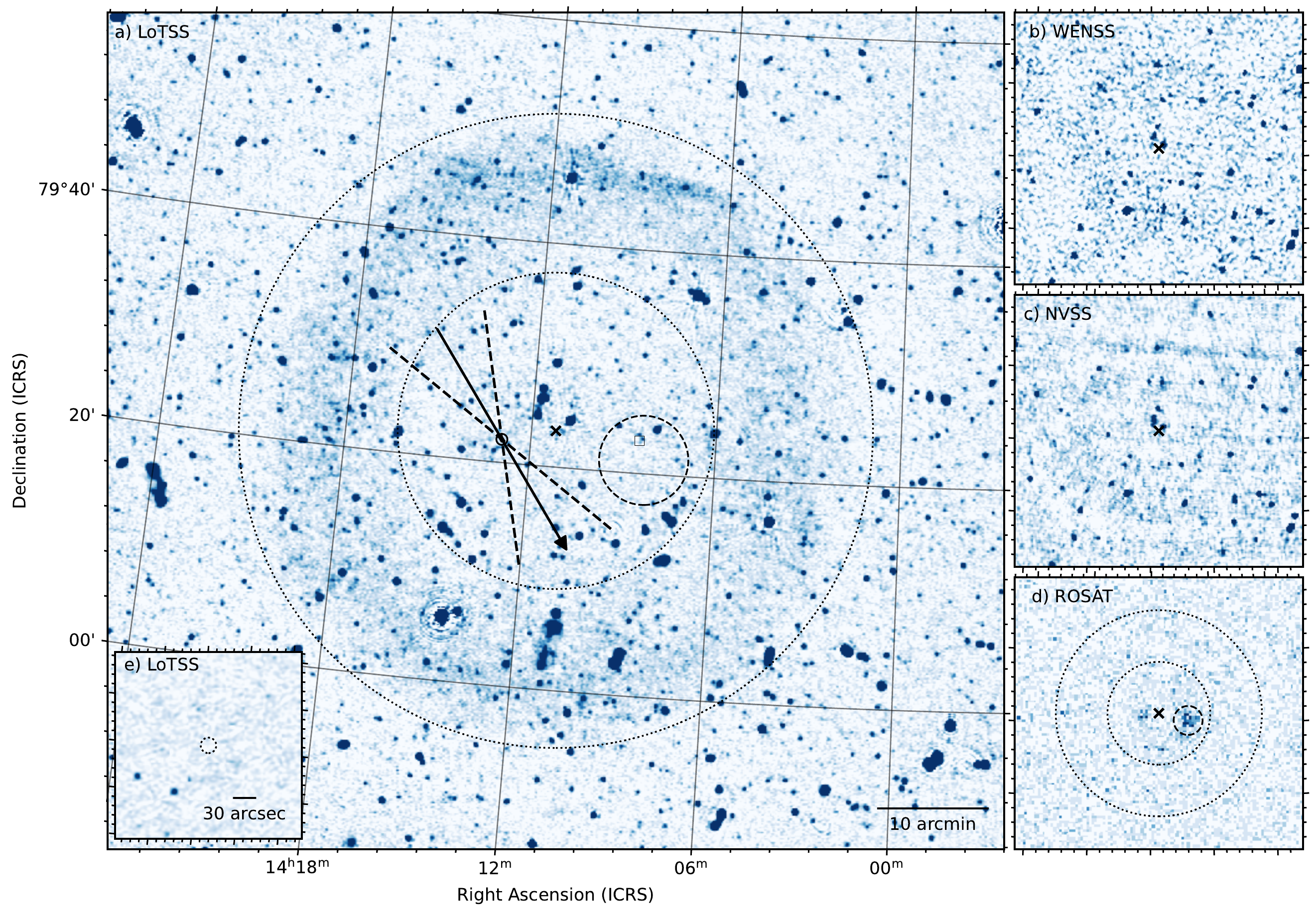}
    \caption{\textit{a)} Radio emission from the diffuse ring at 144\,MHz in the LOFAR HBA LoTSS imaging mosaic ($1\fdg25\times1\fdg25$ field-of-view, $20\arcsec$ resolution). The location of Calvera is indicated with a circle, with the arrow representing the \citet{halpern15} proper motion from 10\,kyr in the past to 10\,kyr into the future. The dashed lines indicate the propagated $1\sigma$ positional uncertainty due to proper motion. The cross sign indicates the estimated center of the diffuse ring, with the dotted circles representing the inner and outer radii. The dashed circle ($4\arcmin$ radius) denotes the location of the diffuse X-ray emission seen by \citet{zane11} from 1RXS\,J140818.1+792113, with the square indicating the location of H$\alpha$ emission. \textit{b)} The diffuse ring in the Westerbork Synthesis Radio Telescope WENSS survey imaging at 325\,MHz and \textit{c)} the Very Large Array NVSS survey at 1.4\,GHz. Both panels have a $1\fdg25\times1\fdg25$ field-of-view and identical pointing and orientation to that of panel \textit{a}. \textit{d)} The 0.1-2.4\,keV X-ray image from the \textit{\textit{ROSAT}} All-Sky Survey \citep{voges99}, with the ring center and inner outer radii of the radio ring denoted by the cross and dotted circles. The dashed circle encircles the emission from 1RXS\,J140818.1+792113. \textit{e)} A $2\arcmin\times2\arcmin$ section of the $6\arcsec$ resolution LoTSS imaging mosaic at the location of Calvera, indicated with the dotted circle ($10\arcsec$ radius).}
    \label{fig:skycharts}
\end{figure*}

\section{Observations and analysis}
\subsection{Radio continuum imaging}
Visual inspection of radio continuum images obtained as part of the ongoing LOFAR Two-metre Sky Survey (LoTSS; \citealt{shimwell17}) led to the discovery of a $\sim1\degr$ diameter diffuse ring-like structure at $l=118\fdg4$, $b=+37\fdg0$ in the 8\,hr LoTSS pointing P207+77 (observed on 2019 September 19). The presence of the ring was confirmed in two nearby LoTSS pointings, P207+80 and P222+80, observed on 2022 January 29 and 2022 February 6, respectively, both with 8\,hr exposures. The LoTSS pointings were offset from the center of the ring by $2\fdg20$ (P207+77), $1\fdg04$ (P207+80) and $1\fdg93$ (P222+80). These pointings were processed by the LOFAR Surveys Key Science Project team, employing the \texttt{ddf-pipeline}\footnote{\url{https://github.com/mhardcastle/ddf-pipeline}} developed to reduce LoTSS observations \citep{shimwell17, shimwell19, shimwell22}, which delivers direction-dependent corrected mosaics at $6\arcsec$ and $20\arcsec$ resolution \citep[see][for more details]{tasse21}. The imaging is done using the multiscale cleaning algorithm described in \cite{offringa17}.

Since the edges of the ring are not sharp, we estimate its center, radius and width by modelling it as a circularly symmetric radially offset Gaussian. Starting with a $1\fdg25\times1\fdg25$ section centered on the ring from the $20\arcsec$ resolution LoTSS image, all pixels above the 98\% percentile (0.7\,mJy\,beam$^{-1}$) were masked. The masked shapes were expanded by one pixel through a single binary dilation operation to mask the fainter wings of point sources. In the resulting image, 3.1\% of the pixels were masked. The unmasked pixels were used in a least-squares minimisation of the circularly symmetric Gaussian, fitting for the Gaussian width, the radial offset around the center, as well as the Gaussian amplitude above some constant background. 
%The fit constrains the ring center to $\alpha=14^\mathrm{h}11^\mathrm{m}12\fs6$, $\delta=+79\degr23\arcmin15\arcsec$, with an estimated uncertainty of $15\arcsec$. The radius and Gaussian width of the ring are $r=21\farcm3$ and $\sigma=3\farcm6$, resulting in a FWHM of $8\farcm5$.

To measure the integrated flux density of the ring, we further processed the LoTSS data using the ``extraction and selfcal'' method described in \citet{vanweeren21}. This methods allows one to correct for residual calibration artifacts while permitting fast and flexible reimaging of a smaller portion of the dataset. We focused on a squared region of 1.1 deg$^2$ centered on the ring and obtained a new image with WSClean \citep{offringa14} at 30\arcsec\ resolution. In this image, obtained using a minimum uv-cut of $60\lambda$ (corresponding to an angular size of $\sim57\arcmin$), discrete sources were subtracted from the visibilities using the clean components created from a previous imaging run at higher resolution (created with an minimum uv-cut of $1000\lambda$ to filter out the ring). 
%The integrated flux density of the radio ring measured at the central frequency of 144 MHz is $1.08\pm0.15$~Jy. The uncertainty includes the statistical error, and the systematic error on the flux scale (10\% of the measured flux density of the ring).

\subsection{Radio Polarisation}
The LoTSS data products include linear polarisation Stokes \textit{Q} and Stokes \textit{U} cubes with a spatial resolution of 20\arcsec. 
To search for linearly polarized emission, we used the rotation measure synthesis technique \citep{brentjens05} on the 3\arcmin\ resolution \textit{Q} and \textit{U} channel images with \texttt{pyrmsynth\_lite}\footnote{\url{https://github.com/sabourke/pyrmsynth_lite}}. There were 480 \textit{Q} and \textit{U} images from 120~MHz to 168~MHz. We searched the band-averaged polarized intensity images across a Faraday depth range of $\pm100$~rad~m$^{-2}$ with a sampling of 0.5 ~rad~m$^{-2}$ and found no evidence for polarized emission associated with the ring, at a level of  300 $\mu$Jy~beam$^{-1}$. This is a higher noise than typical for LoTSS images, but is due to the fact that the ring is located at the edge of the observing field. More details on the rotation measure synthesis
method developed for LoTSS data products can be found in O’Sullivan et al. (in prep). 

\subsection{Beamformed radio observations}
A high-time resolution Stokes \textit{I} observation (ObsID: L257877) of Calvera was obtained on 2015 January 16 using the inner 22 LOFAR HBA core stations in beamforming mode, providing a $3\farcm5$ tied-array beam pointed at the position of Calvera. The observation was 3\,hr in length with 78.125\,MHz of bandwidth, centred at 149\,MHz, and with 163.84\,$\upmu$s time resolution and 12.2\,kHz frequency resolution. For this observational setup, smearing due to dispersion reaches 1\,ms at a dispersion measure (DM) of about 30\,pc\,cm$^{-3}$ and 3\,ms at $\mathrm{DM}=80$\,pc\,cm$^{-3}$. Models for the Galactic electron distribution predict maximum dispersion measures (DMs) of 42.2\,pc\,cm$^{-3}$ at 10.1\,kpc (NE2001; \citealt{cordes02}) and 35.5\,pc\,cm$^{-3}$ at 15.2\,kpc (YMW16, \citealt{yao17})

The observation was folded with the \citet{halpern15} X-ray timing ephemeris using \texttt{dspsr} \citep{vanstraten11} to 10\,s sub-integrations. Though the observation epoch falls outside of the timing span of both the \citet{halpern15} and \citet{mereghetti21} timing ephemerides, extrapolating these to the observation epoch yields consistent folding periods ($\Delta P = -17.6\pm6.9$\,ns). Using \textsc{psrchive} \citep{hotan04} tools, radio frequency interference (RFI) was masked and the folded observation was subsequently dedispersed to 401 trial dispersion measures (DMs) between 0 and 80\,pc\,cm$^{-3}$ at 0.2\,pc\,cm$^{-3}$ steps. Each DM trial was averaged to 128 pulse phase bins, 0.78\,MHz frequency channels and 100\,s sub-integrations and searched for periodicities within 280\,ns around the folding period and DM offsets of $-0.23$ to $0.23$\,pc\,cm$^{-3}$ with \texttt{pdmp}. No periodicities with a significance over $7\sigma$ were found. The flux calibration method by \citet{kondratiev16} estimates sensitivity limits of 0.25\,mJy for an assumed fractional pulse width of 10\% (0.8\,mJy for 50\%) at $7\sigma$.

The observation was also searched for emission from single pulses using tools from the \textsc{presto} software suite \citep{ransom01}. After RFI masking, dedispersed timeseries were created for DMs up to 80\,pc\,cm$^{-3}$, with steps of 0.002\,pc\,cm$^{-2}$ up to 10\,pc\,cm$^{-3}$ at the native time resolution, 0.005\,pc\,cm$^{-3}$ from 10 to 22.5\,pc\,cm$^{-3}$ for a downsampled time resolution of 0.655\,ms, and 0.02\,pc\,cm$^{-3}$ at 1.31\,ms. Single pulse candidates with DMs below 0.2\,pc\,cm$^{-3}$ were discarded due to pollution by zero DM RFI, while all other events above a significance of $7\sigma$ were visually inspected. Unfortunately, no astrophysical pulses were detected as all remaining events could be attributed to time ranges affected by RFI. We estimate $7\sigma$ fluence limits of 2.6\,Jy\,ms for 1\,ms wide signals.

\begin{figure*}
    \centering
    \includegraphics[width=\textwidth]{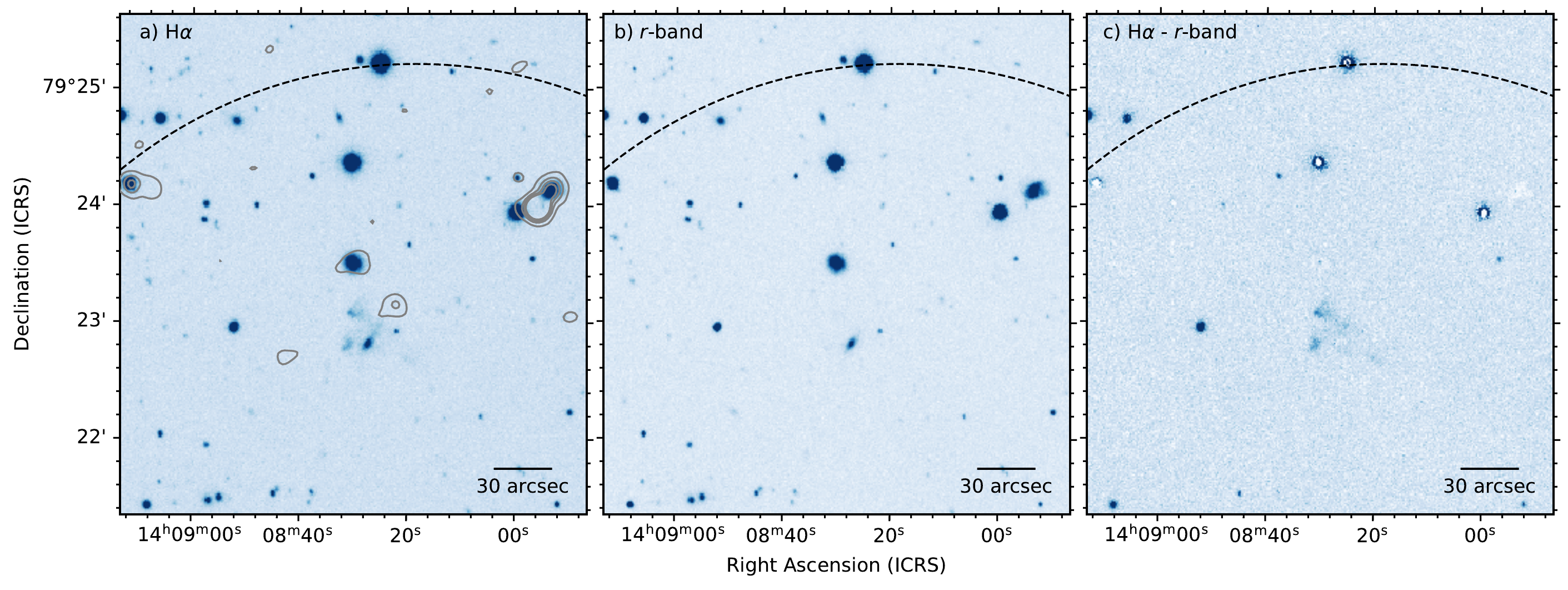}
    \caption{The detection of extended H$\alpha$ emission in the INT/WFC observations in the narrowband H$\alpha$ filter (panel \textit{a}), the broadband $r$-band filter (panel \textit{b}) and the continuum subtracted H$\alpha - r$-band (panel \textit{c}). Each panel shows a $4\arcmin\times4\arcmin$ field-of-view, with the dashed circle representing extent of the diffuse X-ray emission from 1RXS\,J140818.1+792113 (see Fig.\,\ref{fig:skycharts}). Radio contours from the $6\arcsec$ LOFAR mosaic are overlaid in panel \textit{a}.}
    \label{fig:halpha}
\end{figure*}

\subsection{Optical Observations}

We also observed the region around Calvera and a section of the radio ring with the the Wide Field Camera (WFC) on the 2.5m Isaac Newton Telescope (INT) in La Palma. Our initial aims were to search for optical forbidden line emission from the radio shell that would indicate radiative SNR shocks, and the presence of a bow shock near Calvera that could help us constrain its direction of motion. The region was observed for three nights (2022 March 19, 20, and 23) in the H$\alpha$ and $r$-band filters. 

In order to reduce the optical observing data we used the automated astronomical data reduction pipeline from THELI \citep{THELI_Erben_2005,THELI_Schirmer_2013}. THELI provided the astrometry and relative photometry to combine the observations of a single observation night. For every single observation THELI subtracted the sky with a model. To model the sky THELI uses SExtractor \citep{SExtractor_Bertin_1996} to remove all sources, dynamically fills the gaps in the field and convolves the resulting field with a Gaussian kernel of 256 pixels. Finally we used THELI to coadd the observations of a single observation night and perform a flux calibration with known field sources from Pan-STARRS \citep{PANSTARRS_Chambers_2016}. For the $r$-band and H$\alpha$ we also combined the observations for different nights. The $r$-band and H$\alpha$ images were photometrically calibrated against the PanSTARSS $r$-band magnitudes \citep{chambers16} using THELI.

No emission associated with either the radio ring or a bow shock was detected in H$\alpha$ (174 minutes of exposure time). We estimate a $5\sigma$ limit on H$\alpha$ surface density flux of $1.1\times10^{-16}$\,erg\,cm$^{-2}$\,s$^{-1}$\,arcsec$^{-2}$.

We did detect, though, a smudge of H$\alpha$ emission centered at $\alpha_\mathrm{J2000}=14^\mathrm{h}08^\mathrm{m}28\fs3$, $\delta_\mathrm{J2000}=+79\degr22\arcmin54\farcs0$. The extended emission is faint, with an H$\alpha$ surface brightness of 24.0\,mag\,arcsec$^{-2}$, somewhat V-shaped, and at its longest it is $23\arcsec$ across.
Figure \ref{fig:halpha} shows the H$\alpha$ and $r$-band images, as well as the continuum subtracted H$\alpha$ image for a region surrounding the smudge. There is no clear radio emission associated with the H$\alpha$ smudge, although there is a point source that almost overlaps with the smudge, and so it is hard to tell if faint extended emission is also present. We examined the four \textit{WISE} bands for emission in this region and found no clear IR counterpart. The smudge is unlikely to be of extra-galactic origin for several reasons: its angular scale, the fact that it does not overlap in full with a galaxy, and the fact that it falls within the redshift range of the filter ($z<0.007$, for an H$\alpha$ filter bandpass of 95~\AA\footnote{\url{https://www.ing.iac.es/astronomy/instruments/wfc/}}).

\section{Results}

\subsection{Detection of a radio ring}
We report the detection of radio emission from a degree-scale diffuse ring in LoTSS radio continuum imaging at 144\,MHz. Figure\,\ref{fig:skycharts}a shows the diffuse ring in the LoTSS radio continuum mosaic. The ring is located at $\alpha=14^\mathrm{h}11^\mathrm{m}12\fs6$, $\delta=+79\degr23\arcmin15\arcsec$, corresponding to a Galactic position of $l=118\fdg41$, $b=37\fdg03$. The ring has a FWHM of $8\farcm5$ and we conservatively estimate an the inner diameter of $28\farcm4$ and an outer diameter of $56\farcm8$ (see Fig.\,\ref{fig:skycharts}). These correspond to a physical inner and outer diameter of 8.2\,pc and 16.5\,pc for a distance of 1\,kpc. The ring does not seem to have a sharp inner or outer edge, and its brightness is relatively constant, both radially and in position angle, except for a brighter, straight filament in the North-West. After subtracting the point sources from the LoTSS mosaic, we measure an integrated flux density of $1.08\pm0.15$~Jy at 144\,MHz (corresponding to the area between the dotted concentric circles in Fig. \ref{fig:skycharts}). The uncertainty includes the statistical error, and the systematic error on the flux scale \cite[10\% of the measured flux density of the ring, following LoTSS Data Release 2, ][]{shimwell22}.

{\def\arraystretch{1.5}\tabcolsep=5pt
\begin{table}[]
\begin{tabular}{cccc}
\textbf{Survey} & \textbf{\begin{tabular}[c]{@{}c@{}}Frequency\\ (MHz)\end{tabular}} & \textbf{\begin{tabular}[c]{@{}c@{}}Flux density\\ (Jy)\end{tabular}} & \textbf{\begin{tabular}[c]{@{}c@{}}Error in flux \\ density (Jy)\end{tabular}} \\ \hline
LoTSS           & 144                                                                & 1.08                                                                 & 0.15                                                                           \\
WENSS           & 325                                                                & 0.63                                                                 & 0.21                                                                          \\
NVSS            & 1420                                                               & 0.22                                                                & 0.04                                                                  
\end{tabular}
\caption{Integrated flux density of the ring centered at $\alpha=14\mathrm{h}11\mathrm{m}12\fs6$, $\delta=+79\degr23\arcmin15\arcsec$, with inner and outer radii of $14\farcm2$ and $28\farcm4$ at various frequencies.}
\end{table}
\label{table:1}
}

Knowing the location and extent of the ring, we searched for its presence in other radio continuum surveys. Radio emission of the diffuse ring is detected at low significance in the Westerbork Northern Sky Survey \cite[WENSS,][]{rengelink97} at 325~MHz (Fig.\,\ref{fig:skycharts}b), and at even lower significance in the NRAO VLA Sky Survey \cite[NVSS,][]{condon98} at 1.4~GHz (Fig.\,\ref{fig:skycharts}c). The ring is not detected in the NRAO VLA Low-frequency Sky Survey (VLSS; \citealt{cohen07,lane14}) at 74\,MHz (0.7\,Jy\,beam$^{-1}$ sensitivity) and the TIFR GMRT Sky Survey (TGSS; \citealt{intema17}) at 150\,MHz (5\,mJy\,beam$^{-1}$ sensitivity).
%The ring is also detected in the LOFAR LBA Sky Survey \cite[LoLSS,][this region also unpublished]{degasperin21}.
The integrated flux densities of the ring in the surveys where it is detected are summarised in table \ref{table:1}, and correspond to a best-fit spectral index value of $-0.71\pm0.09$ (for power-law radio spectra with observing frequency $\nu$ of the form $S_\nu\propto\nu^{\alpha}$).

The full area encompassed by the radio ring has only been observed in the X-rays as part of the \textit{\textit{ROSAT}} All-Sky Survey \citep{voges99}\footnote{It has likely also been observed by eROSITA \citep{sunyaev21}, but the observations are not public as of this writing.}, and no X-ray emission coincident with the radio emission is detected (see Fig.\,\ref{fig:skycharts}d). Two \textit{ROSAT} X-ray sources from the bright source catalog \citep{voges99} are coincident with the extent of the ring, the Calvera X-ray pulsar (1RXS\,J141256.0+792204) and 1RXS\,J140818.1+792113. \citet{zane11} find that the latter source is extended and has a soft spectrum in \textit{XMM-Newton} observations covering the inner region of the ring, and that its X-ray spectrum best modelled as a non-equilibrium ionization (NEI) plasma.

The Calvera X-ray pulsar is offset by $4\farcm9$ from the center of the radio ring, well within its inner radius. No other radio, X-ray or gamma-ray pulsar from the ATNF Pulsar Catalogue (\citealt{manchester05}, version 1.67) is located within the ring. Another 8 radio pulsars are within a $10\degr$ radius, with the closest at an offset of $2\fdg85$.

\subsection{Non-detection of Calvera in LOFAR observations}
No radio emission is detected in the LoTSS radio continuum mosaic at the location of Calvera (Fig.\,\ref{fig:skycharts}e), either as a point source from pulsed or unpulsed emission, or as extended emission from a pulsar wind nebula (PWN). The rms noise measured at the location of Calvera in the 6\arcsec\ resolution map is 75~$\mu$Jy~beam$^{-1}$; therefore, we estimate a $3\sigma$ flux density upper limit of 225\,$\mu$Jy for Calvera's radio continuum emission at 144\,MHz. 

The beamformed LOFAR observations also fail to detect pulsed radio emission at the known spin period of Calvera, or as single pulses down to flux density and fluence limits of 0.25\,mJy and 2.6\,Jy\,ms at 149\,MHz, respectively. These limits add to previous pulsed flux density upper limits of 4\,mJy at 385\,MHz and 0.3\,mJy at 1380\,MHz by \citet{hessels07} using the Westerbork Synthesis Radio Telescope and 0.05\,mJy at 1360\,MHz with the Effelsberg telescope \citep{zane11}. These observations confirm that Calvera is a radio quiet pulsar.

\subsection{Statistical modelling of the probability of chance alignment}
\label{sec:modelling}

Although chance alignments between pulsars and SNRs are frequent in the Galactic plane \citep{gaensler95}, at such high Galactic latitudes both pulsars and remnants are rare. We performed some Monte Carlo simulations so as to quantify the likelihood of a chance alignment. 

In order to find the probability that a pulsar falls within a fixed area with the coordinates and diameter of our radio ring (which we take to be one degree), we simulate many iterations of pulsar populations and count the number of times that a pulsar falls within the radio circle.
Each sample pulsar is generated with a set of Galactocentric coordinates $(r,\theta,h)$, where $r$ is the Galactocentric radius, $\theta$ is the polar angle, and $h$ is the height above or below the Galactic plane. We sample $r$ according to the distribution given by \cite{yusifov04} in Galactocentric radius, and $\theta$ by drawing from a uniform distribution between 0 and $2\pi$. 
To sample $h$, we tried using exponential distributions with scale heights of 100~pc, 330~pc and 350~pc \citep{faucher06,lorimer06,mdzinarishvili04} and found that the simulated sample, once we converted to heliocentric coordinates, was very compact in Galactic latitudes. In runs of 10,000 Monte Carlo iterations, the rate at which a pulsar fell within the area of the radio ring was consistently less than $5\times10^{-4}$. This distribution produced on average 1 pulsar within a 10$\degr$ radius centered at Calvera, whereas we observe 9 in such a region, which suggests that the scale height distributions in the above references do not describe well this local region of the Galaxy.

Given that the simulated populations sampled from the exponential distributions seemed to generate pulsars whose Galactic latitudes were too low, we also simulated the distribution in $h$ empirically. We used the ATNF Pulsar Catalogue v1.67 \citep{manchester05}, removing pulsars in globular clusters or in the Magellanic Clouds, in order to histogram the distribution in heights above the Galactic plane (using the dispersion measure distances). Then, we sampled from the midpoints of the histogram bins, adding some noise so as to not sample in discrete heights above or bellow the plane. Naturally, since we sampled from the ATNF Pulsar Catalogue v1.67 empirical distribution, we recover the observed value of 9 pulsars in a 10$\degr$ radius on average. Our Monte Carlo simulations gave a probability of chance alignment smaller than 0.01.

These calculations are back-of-the-envelope and cannot be the basis to claim that Calvera and the radio ring are aligned, but they are a good indication that it is very unlikely that their positional coincidence is by chance. 

\section{The nature of the radio ring}

\subsection{An H II region}

H II regions have a radio spectral index  of $>-0.2$ \citep{condon16}, which is already quite different from the best-fit spectral index value of $-0.71\pm0.09$ that we find here. H~II regions also emit profusely in H$\alpha$ \cite[e.g.,][]{haffner09}, which we do not detect. Moreover,
the area around Calvera was observed as part of the Wide-field Infrared Survey Explorer \cite[\textit{WISE},][]{wright10} all-sky survey. 
We examined the four \textit{WISE} bands for emission coincident with the radio ring and detected no emission above the image noise.
\cite{anderson14} argue that the \textit{WISE} survey should be able to
detect the mid-infrared emission from \textit{all} Galactic H II regions. Moreover,
\cite{makai17} calculated the correlations between the radio and IR flux densities of Galactic H~II regions. Using the values (for regions $>1$~pc) from their table 2, and a radio spectral index of $-0.1$, we would expect $F_{8~\mu\mathrm{m}} \sim 40$~Jy, $F_{12~\mu\mathrm{m}} \sim 80$~Jy and $F_{22~\mu\mathrm{m}} \sim 90$~Jy if the ring were an H~II region. Even for a source with an area $>1,800$ arcmin$^2$, these values are well within the extended source sensitivities of their respective \textit{WISE} bands. 

In addition to the electromagnetic evidence against the H~II interpretation, the morphology also points against it. The circular shape of the source seems to imply that it is ionised from within; however, a query of a 1 degree radius around the coordinates of Calvera in the Catalog of Galactic OB Stars \citep{reed03} yielded no results.
We therefore discard the possibility that the ring is an H~II region.

\subsection{A supernova remnant}

\begin{figure}
    \centering
    \includegraphics[width=\columnwidth]{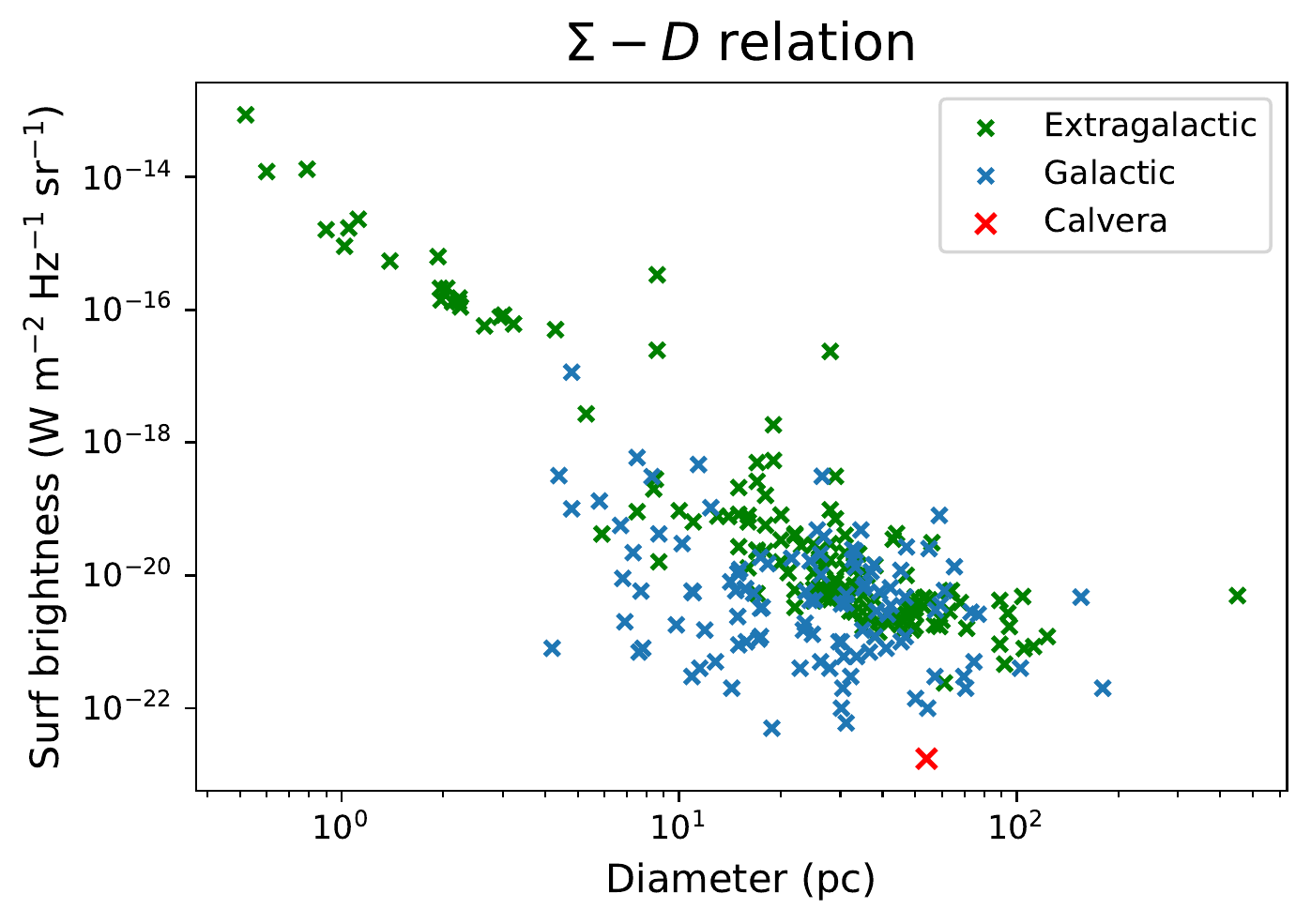}
    \caption{1~GHz surface brightness versus diameter for the \cite{vukotic19} sample of Galactic SNRs with known distances (blue) and the \cite{urosevic05} sample of extragalactic SNRs. The purported Calvera SNR is plotted in red using the measured surface brightness at 144~MHz and the best-fit spectral index of 0.71, and the 54~pc diameter derived from the 3.3~kpc distance estimate of \cite{mereghetti21}.}
    \label{fig:sigd}
\end{figure}

The measured radio spectral index of $-0.71\pm0.09$ already presents a good indication that the radio emission is of a synchrotron origin, characteristic of SNRs \citep{dubner15}. In addition to this, there are two main pieces of evidence that support a SNR interpretation for the ring around Calvera: the low probability of a chance alignment between the pulsar and the radio source (see section \ref{sec:modelling}), and the presence of the extended X-ray source 1RXS J140818.1+792113.
The fact that a possible association exists between all three objects is what strengthens the SNR interpretation. Of course, there still exists the possibility that the source is a SNR with no relation to the Calvera pulsar other than the positional overlap. 
However, this would require two stars exploding in the same line-of-sight, at a Galactic latitude where SNRs are infrequent \cite[the][catalogue lists 294 SNRs, only three of which are at $b>10\degr$, and none are above $b = 15\degr$]{green19}. Our favoured interpretation is that this radio ring is the remnant of the supernova explosion that formed the Calvera pulsar. This would make the source SNR G118.4+37.0, or Calvera's SNR. 

Perhaps the most convincing piece of evidence that the radio emission is from a SNR is 1RXS J140818.1+792113: \cite{zane11} find extended soft X-ray emission in \textit{XMM-Newton} MOS and \textit{\textit{ROSAT}} All-Sky Survey (RASS) maps (indicated with a dashed line in Figs. \ref{fig:skycharts}a and \ref{fig:skycharts}d).
They perform a spectral analysis on the data and find that the spectrum is well-fitted by a non-equilibrium ionisation (NEI) model
with an overabundance of oxygen; the former is a
typical signature of SNRs in general, and the latter of SNRs with a core-collapse origin \citep{vink12}. The extended emission is also visible in the public \textit{Chandra} HRC-I observations used in \cite{halpern15}.
The region of extended emission has angular separation of 13\arcmin\ with Calvera, which lead \cite{zane11} to suggest that the two are unrelated. However, both are encompassed by the ring, and all three (the pulsar, the NEI plasma, and the shell) could well be the remains of a single stellar explosion. The additional possibility that only two of these components are related -- the plasma and the ring but not the pulsar, or any such combination -- exists. However, the presence of a SNR plasma inside the radio ring is very indicative that the ring is a SNR, irrespective of whether the two are associated with Calvera. A final point to be made regarding the NEI plasma is that the relation between the H$\alpha$ smudge and the X-ray emitting material remains unclear. Follow-up spectroscopy of the smudge could determine whether both have similar compositions.

\cite{mereghetti21} estimated the distance to the Calvera pulsar to be 3.3~kpc (by assuming a neutron star radius of 13~km, and fitting the thermal emission), and that the supernova explosion occurred in the Galactic halo. At this distance, the SNR would have a diameter of 54~pc, and it would be at a height over the Galactic plane of 2~kpc. This would make Calvera's SNR, along with the remnant of SN1006 \citep{raymond17}, one of the known remnants with a largest height above the Galactic plane (with the important difference that Calvera's SNR would be the remnant of a core-collapse explosion). Getting the (massive) progenitor star to such a large height above the Galactic plane could require a high velocity runaway scenario: a star with a vertical velocity of 100~km~s$^{-1}$ would take $\sim30$~Myrs to reach it, which points to a star in the low-mass end of core-collapse explosions ($8-10~\msun$), or alternatively a star in some kind of binary interaction that extends its lifetime \cite[e.g., as discussed in][]{zapartas21}. Hypervelocity runaway stars are rare \citep{renzo19} but certainly can be responsible for core-collapse supernovae in the outskirts of galaxies \cite[e.g.,][]{zinn11}.
Alternatively, the neutron star need not be formed by a massive star at all, if its origin is the accretion-induced or merger-induced collapse of a white dwarf \citep{grindlay88, saio85}. This solves the problem of the location of the progenitor, although the oxygen overabundance noted in the X-ray plasma by \cite{zane11} favours a core-collapse origin.

For SNRs, there exists a relationship between their radio surface brightness at 1~GHz and their diameters, known as the $\Sigma-D$ relation. Some authors \cite[e.g.,][]{green04} are sceptical of the soundness of this relation; however, when we plot the diameter as derived from the \cite{mereghetti21} distance against the radio surface brightness for the purported Calvera SNR (see Fig. \ref{fig:sigd}), the data point falls roughly in line with the samples of Galactic and extragalactic SNRs found in \cite{vukotic19} and \cite{urosevic05}, respectively. Using the values for the $\Sigma-D$ relation as found in \cite{vukotic19}, for a 1~GHz surface brightness of $1.74\times10^{-23}$~W~m$^2$~Hz$^{-1}$~sr$^{-1}$ (using the best-fit spectral index of $-0.71$ and the area between the inner and outer radii), we find a diameter of 62~pc, quite comparable to the 54~pc diameter as derived from the \cite{mereghetti21} distance.

An X-ray proper motion measurement by \citet{halpern15} places the location of Calvera at the time of explosion $\sim20$\arcmin\ away from the center of the ring (for an explosion that happened 8,000 years ago, see discussion below; for an explosion corresponding to the characteristic age of the pulsar, this distance is $>6$\degr). This could be interpreted as an indication that the two are not related. However, the uncertainties in the \textit{Chandra} observations that produced this result (dominated by the short 2\,ks HRC-I exposure obtained in 2007), and the uncertainty in the age of the pulsar itself, are large. Moreover, an inhomogeneous environment can easily produce an offset between the SN explosion site and the SNR's geometric center \citep{dohm-palmer96}. This need not result in an asymmetric remnant: \cite{williams13} showed that a uniform explosion into a density gradient produces a circular remnant, but the center of the circle is different from the explosion site. These are plausible reason for the observed misalignment; we cannot rule out association simply from these proper motion measurements. 

Another issue that draws attention to itself is that the X-ray plasma does not overlap with the radio emission. This is uncommon, but not unheard of, in SNRs: for instance, RCW 86 has a patch of X-ray emission with no radio counterpart \citep{bamba00}. The presence of the optical smudge is also puzzling: the only interpretation we can think of is that the SNR shock encountered a small density discontinuity in its surrounding medium that caused it to become radiative at that location.
\iffalse
SNRs with thermally emitting X-rays \textit{internal} to a radio synchrotron shell are known as mixed-morphology SNRs \cite[MM SNRs,][]{rho98}, and their morphology is typically explained by invoking a dense medium \cite[see discussion in][ and references therein]{vink12}. If the ring is indeed the remnant of an explosion that occurred in the halo, then in order to account for its mixed-morphology we would have to invoke aspects of the progenitor history (e.g., a bow shock, a stellar wind) that could have triggered an early reverse shock. This is at odds with the fact that the remnant is so circular that its environment must be relatively homogeneous.
\fi

Another source of tension comes from the pulsar characteristic age, $\tau_\mathrm{c}=2.9\times10^{5}$\,yr \citep{halpern13}, which is well past the time when we expect an SNR to still be visible. However, characteristic ages often overestimate the true age of the pulsar; this is due to the fact that the characteristic age neglects the initial period of the pulsar.  Since 59 ms is well within the distribution of birth periods inferred from modeling the pulsar population \cite[e.g.][]{faucher06}
Calvera's characteristic age is only an upper limit to its true age. A case on point are central compact objects (CCOs): thermal
X-ray sources with no accompanying radio or gamma-ray emission that are associated with known SNRs, and which have characteristic ages $\sim10^8$~yrs \citep{deluca17}. Moreover, although in general SNRs live for $\sim50,000$~years, we do not know what is the spread in lifetimes --it could be intrinsically large, given that a source's lifetime is likely highly dependent on its environmental
conditions. An SNR in the halo, where the ambient densities are low, could well live longer than 50,000~years. However, the Sedov-Taylor self-similar solution \citep{sedov59,taylor50}, for an explosion energy of 10$^{51}$~erg and a SNR radius of 54~pc gives a SNR age much younger than that: 
$7,700$~years for an ambient density of $4\times10^{-4}$~cm$^{-3}$ \cite[we estimated the ambient density in the halo around Calvera using the model of][with $r=9.5$~kpc the distance between Calvera and the center of the Galaxy] {miller13}. We do not know whether the conditions for self-similar evolution are present on this remnant, though, so this age estimate, too, needs to be taken with a grain of salt.

Finally, the ring shows no clear optical line emission. Abundant optical line emission is a common feature in aged SNRs: when a SNR shock is slowed down to approximately 200~km~s$^{-1}$ it begins to emit profusely in optical forbidden lines, and cools quickly \citep{vink12}. However, a SNR's cooling time and length scales depend on the pre-shock density \citep{raymond79}, and in low density environments the SNR enters the radiative phase at lower velocities (corresponding to older ages). It is possible that the optical line emission is too faint for our optical observations to recover, or that the remnant still has not reached its radiative phase. 
 
\subsection{An Odd Radio Circle}

Odd Radio Circles, or ORCs \citep{norris21a}, are a recently discovered class of diffuse extragalactic objects,
first detected in the Pilot Survey of the Evolutionary Map of the Universe \citep{norris21c}.
They are faint, arcminute-sized circles of steep-spectrum radio emission at high Galactic latitudes with no corresponding diffuse 
emission at optical, infrared, or X-ray wavelengths \citep{norris21b}. As of this writing, there are only seven published ORCs \citep{norris21b,koribalski21,filipovic22,omar22},
three of which have an elliptical galaxy at their centre (referred to as their \lq host\rq\ galaxy). 
It is unclear what phenomena produce ORCs, and there is no evidence that the radio shells are associated 
with their \lq host\rq\ galaxies other than the positional coincidence of the two. In fact,
the most recently reported ORC \citep{filipovic22} was suggested to be the remnant of an exploded runaway star that left its galaxy (the Large Magellanic Cloud),
whereas the remaining ones are thought to be distant ($z>0.1$) objects. %Moreover, the Sedov-Taylor solution for self-similar SNR evolution \citep{sedov59,taylor50} is also dependent on the ambient density

If further work disproves that the ring is a SNR, then it could be LOFAR's second contribution to the list of known ORCs \cite[the first being][]{omar22}. The Pan-STARRS PS1 images and catalog \citep{chambers16} shows no prominent galaxy near the center of the ring as seen in other ORCs; out of the 18 PS1 objects located within $1\arcmin$ from the center of the ring, only a single source (WISE\,J141113.52+792336.4) appears slightly extended ($1\arcsec\times2\arcsec$) and has WISE colors consistent with those of spiral galaxies. It is unclear whether this could be the ORC host galaxy if the radio ring would be identified as an ORC. However, the known ORCs are arcminute-sized objects, whereas the ring is $\sim1$ degree in diameter. If the ring is an extra-galactic object with comparable redshift to the other ORCs, this would require that the mechanism that produces the ORCs is able to create structures that span 1.5 orders of magnitude. A more likely scenario is that this is an ORC of the type found by \cite{filipovic22}: a nearby source, rather than a distant object. A nearer location would solve the issue of its large scale but would take us back to the SNR interpretation. A final point is that, morphologically, the ORCs appear to be quite different from the ring discussed in this paper, as they tend to have more structure and be less symmetric.

\section{Discussion}

We are fairly confident that this ring of radio emission seen in LOFAR, WENSS, and the NVSS is the remnant of the explosion that formed Calvera. The main pieces of evidence are the presence of an X-ray plasma with enhanced metals and the rarity of a chance alignment with a pulsar at such high Galactic latitudes. We discard the H~II region interpretation given the lack of infrared and H$\alpha$ emission; the ORC interpretation remains a possibility, but the difference in size with the other known ORCs and the lack of a clear host galaxy candidate weaken it. An additional possibility is that the ring is a SNR unrelated to Calvera, although this would require two supernova explosions occurring in a single line-of-sight a high Galactic latitude, which is unlikely.

More data are necessary to determine whether the ring is an SNR or an ORC. In particular, X-ray spectroscopy of the brighter filament at the North West could unambiguously settle whether the ring is indeed a SNR. We do not know if the ring shows no optical emission at all, or whether the optical emission is too faint for our INT observations to detect, but an optical follow-up of the ring will require deep exposures on dark conditions. Another set of follow-up observations that could be more interesting is optical spectroscopy of the H$\alpha$ smudge, to clarify whether metals are also present in its spectrum, and whether it has a similar composition to the X-ray emitting plasma. Finally, LOFAR LBA observations of the area at 58~MHz will better constrain the radio spectral index, and  polarisation studies at GHz frequencies can also help settle whether the ring is a SNR.

There are some implications for the Calvera pulsar if the radio ring is a SNR associated with it. Several authors \citep{zane11,halpern11,halpern13} have suggested that Calvera might be an \lq orphaned\rq\ or aged CCO, whose associated SNR either had faded away or was too faint to detect. The discovery of a SNR around it would place Calvera in the CCO category. The age of the pulsar would almost certainly be different from its characteristic age, since the remnant seems to be on the younger side.

If confirmed to be an SNR, the ring around Calvera would be one of few SNRs with large heights above the Galactic plane. This would make it an interesting object for studies of SNR evolution in diffuse environments.
Moreover, SNRs shock and heat the material they expand into, and in this way make visible the medium around them. SNRs at high Galactic latitudes are in a unique position to probe the interstellar medium (ISM) of the Milky Way halo. The combined analysis of deep X-ray and radio surveys, such as the one conducted by the \textit{eROSITA} telescope \citep{predehl21} onboard the \textit{SRG} observatory \citep{sunyaev21}, and the suite of surveys carried out with MeerKAT and LOFAR could discover a population of low surface brightness, high Galactic latitude remnants \cite[e.g.,][or the Calvera SNR in this work]{churazov21,churazov22,becker21}. These remnants could be longer-lived and fainter than SNRs in the Milky Way disk, given the low ambient densities in the halo.

\section{Summary}

   \begin{enumerate}
       \item We report the LOFAR 144~MHz discovery of a ring of diffuse radio emission centered at $\alpha=14\mathrm{h}11\mathrm{m}12\fs6$, $\delta=+79\degr23\arcmin15\arcsec$,  with an outer radius $R_\mathrm{out} = 29\arcmin$, an inner radius $R_\mathrm{inn} = 14\arcmin$, and an integrated flux density of $1.08\pm0.15$~Jy. The ring center is offset 4.87\arcmin\ from the location of the Calvera pulsar. The ring shows low-significance emission at 325~MHz and 1.4~GHz, and no emission in either of the four \textit{WISE} bands. Its spectral index is $-0.71\pm0.09$.
       \item We did not detect radio emission at 144~MHz from the Calvera pulsar; we report a 225~$\mu$Jy 3$\sigma$ upper limit for Calvera's flux density at 144~MHz. We also fail to detect pulsed radio emission at the known spin period of Calvera, or as single pulses down to flux density and fluence limits of 0.25 mJy and 2.6 Jy ms at 149 MHz, respectively.
       \item We detect no radiative emission from the ring in H$\alpha$. We do detect a smudge of extended H$\alpha$ emission internal to the ring.
      \item The non-detection of the ring in \textit{WISE}, and its radio spectral index value allow us to discard the possibility that the ring is an H~II region.
      \item We estimate the probability of chance alignment between the pulsar and the radio ring to be $<1$\%.
      \item We favour an interpretation whereby the ring is the remnant of the supernova explosion that formed Calvera, although more evidence is needed to support this scenario.
      \item A final possibility is that the ring is an Odd Radio Circle, but this is disfavoured due to the ring's difference in size with other known ORCs.
      \item If the ring is an SNR, it is uniquely suited to probe the ISM in the Galactic halo.
   \end{enumerate}

\begin{acknowledgements}
LOFAR \citep{vanhaarlem13} is the LOw Frequency ARray designed and constructed by ASTRON. It has observing, data processing, and data storage facilities in several countries, which are owned by various parties (each with their own funding sources), and are collectively operated by the ILT foundation under a joint scientific policy. The ILT resources have benefitted from the following recent major funding sources: CNRS-INSU, Observatoire de Paris and Universit\'{e} d'Orl\'{e}ans, France; BMBF, MIWF-NRW, MPG, Germany; Science Foundation Ireland (SFI), Department of Business, Enterprise and Innovation (DBEI), Ireland; NWO, The Netherlands; The Science and Technology Facilities Council, UK; Ministry of Science and Higher Education, Poland; Istituto Nazionale di Astrofisica (INAF), Italy.
This research made use of the Dutch national e-infrastructure with support of the SURF Cooperative (e-infra 180169) and the LOFAR e-infra group. The J\"ulich LOFAR Long Term Archive and the German LOFAR network are both coordinated and operated by the J\"ulich Supercomputing Centre (JSC), and computing resources on the supercomputer JUWELS at JSC were provided by the Gauss Centre for Supercomputing e.V. (grant CHTB00) through the John von Neumann Institute for Computing (NIC). This research made use of the University of Hertfordshire high-performance computing facility and the LOFAR-UK computing facility located at the University of Hertfordshire and supported by STFC [ST/P000096/1], and of the Italian LOFAR IT computing infrastructure supported and operated by INAF, and by the Physics Department of Turin University (under an agreement with Consorzio Interuniversitario per la Fisica Spaziale) at the C3S Supercomputing Centre, Italy. 
The Isaac Newton Telesccope is operated on the island of La Palma by the Isaac Newton Group of Telescopes in the Spanish Observatorio del Roque de los Muchachos of the Instituto de Astrofísica de Canarias. The WFC imaging was obtained as part of the Leiden education program. MA acknowledges support from the VENI research programme with project number 202.143, which is financed by the Netherlands Organisation for Scientific Research (NWO). AB acknowledges support from the VIDI research programme with project number 639.042.729, which is financed by the Netherlands Organisation for Scientific Research (NWO). RJvW acknowledges support from the ERC Starting Grant ClusterWeb 804208. MJH acknowledges support from the UK Science and Technology Facilities Council [ST/V000624/1]. 
\end{acknowledgements}

\bibliographystyle{aa}
\bibliography{biblio.bib}

\end{document}